\documentclass[fleqn,usenatbib]{mnras}

\usepackage[T1]{fontenc}
\usepackage{ae,aecompl}
\usepackage{txfonts}
\usepackage{mathptmx}
\usepackage{graphicx}
\usepackage{amssymb}
\usepackage{aas_macros}
\usepackage{color}
\usepackage{hyperref}

\DeclareMathAlphabet{\pazocal}{OMS}{zplm}{m}{n}

\newcommand{\comment}[1]{} 
\newcommand*{\satellite}[1]{\textit{#1}}

\newcommand*{\hubble}{\satellite{Hubble Space Telescope}}
\newcommand*{\chandra}{\satellite{Chandra}}
\def\clustername{{Cl\,0218.3--0510}}

\title[AGN excess in a high-z protocluster]{Enhancement of AGN in a protocluster at z=1.6}
\author[Charutha Krishnan]{Charutha Krishnan$^{1}$\thanks{E-mail: charutha.krishnan@nottingham.ac.uk}, Nina A. Hatch$^{1}$, Omar Almaini$^{1}$, Dale Kocevski$^{2}$,  
\newauthor Elizabeth A. Cooke$^{3}$, William G. Hartley$^{4}$, Guenther Hasinger$^{5}$, David T. Maltby$^{1}$,  
\newauthor Stuart I. Muldrew$^{6}$, Chris Simpson$^{7}$\\
$^{1}$School of Physics and Astronomy, University of Nottingham, University Park, Nottingham, NG7 2RD, United Kingdom \\
$^{2}$Department of Physics and Astronomy, Colby College, Waterville, ME 04901, USA \\ 
$^{3}$Centre for Extragalactic Astronomy, Department of Physics, Durham University, South Road, Durham, DH1 3LE, United Kingdom \\ 
$^{4}$Department of Physics and Astronomy, University College London, 3rd Floor, 132 Hampstead Road, London, NW1 2PS, United Kingdom \\ 
$^{5}$Institute for Astronomy, University of Hawaii, 2680 Woodlawn Dr, Honolulu, HI 96822, USA \\
$^{6}$Department of Physics and Astronomy, University of Leicester, University Road, Leicester, LE1 7RH, United Kingdom \\ 
$^{7}$Gemini Observatory, Northern Operations Center, 670 N. A`\={o}h\={o}ku Place, Hilo, HI 96720, USA}
\date{Accepted 2017 May 24; in original form 2017 Mar 10}
\pubyear{2017}

\begin{document}
\label{firstpage}
\pagerange{\pageref{firstpage}--\pageref{lastpage}}
\maketitle

\begin{abstract}
We investigate the prevalence of AGN in the high-redshift protocluster \clustername\ at $z=1.62$. Using imaging from the \chandra\ X-ray Telescope, we find a large overdensity of AGN in the protocluster; a factor of $23\pm9$ times the field density of AGN. Only half of this AGN overdensity is due to the overdensity of massive galaxies in the protocluster (a factor of $11\pm2$), as we find that $17^{+6}_{-5}\%$ of massive galaxies ($M_* > 10^{10}$\,M$_{\odot}$) in the protocluster host an X-ray luminous AGN, compared to $8\pm1\%$ in the field. This corresponds to an enhancement of AGN activity in massive protocluster galaxies by a factor of $2.1\pm0.7$ at $1.6\sigma$ significance. We also find that the AGN overdensity is centrally concentrated, located within 3 arcmin and most pronounced within 1 arcmin of the centre of the protocluster. Our results confirm that there is a reversal in the local anti-correlation between galaxy density and AGN activity, so there is an enhancement of AGN in high-redshift protoclusters. We compare the properties of AGN in the protocluster to the field and find no significant differences in the distributions of their stellar mass, X-ray luminosity, or hardness ratio. We therefore suggest that triggering mechanisms are similar in both environments, and that the mechanisms simply occur more frequently in denser environments. 
\end{abstract}



\section{Introduction}
\label{sec:intro}

There is plenty of evidence supporting a correlation between the growth of supermassive black holes (SMBHs) and the formation of their host galaxies. For instance, there is a well known $M$--$\sigma$ relation in the local Universe, correlating the masses of SMBHs and the velocity dispersions of their host galaxies \citep[e.g.,][]{Ferrarese_2000,Gebhardt_2000,Tremaine_2002}. There is also a similar rate of evolution in the emissivity from Active Galactic Nuclei (AGN) and star formation from $z\sim2$ to $z\sim0$ \citep[e.g.,][]{Boyle_1998, Franceschini_1999, Silverman_2008}, implying a link between the accretion of matter onto the SMBH and the build-up of galaxy mass through star formation.

In addition to the correlation between SMBHs and host galaxies, there is also a connection between AGN activity and larger-scale environment. In the local Universe, it has been well established that star formation and AGN activity are suppressed in galaxy clusters. \citet{Dressler_1985} found that the AGN fraction in local massive field galaxies is 5\%, while only 1\% of local cluster galaxies show such nuclear activity. More recently, \citet{Kauffmann_2004} found that twice as many galaxies host AGN with strong [OIII] emission in low-density regions as in high-density regions. This anti-correlation between galaxy density and AGN activity in the local Universe mimics the anti-correlation between galaxy density and the fraction of star forming galaxies. In dense environments, there are several physical processes that could affect the rate of accretion onto the SMBH. Both the availability of cold gas and the mechanisms that funnel the gas into black holes may differ between a galaxy cluster and the field. For instance, in the cluster environment, gas may be removed through environmental processes such as ram-pressure stripping \citep{Gunn_Gott_1972}, and tidal effects due to the cluster potential \citep{Farouki_Shapiro_1981} and other galaxies \citep{Richstone_1976}. These processes, as well as the absence of new infall of cold gas \citep{Larson_1980}, could lead to a shortage of cold gas reservoirs \citep{Giovanelli_1985}, resulting in the suppression of AGN activity \citep{Kauffmann_2004}, reduced star formation activity \citep{Gisler_1978}, and the abundance of post-starburst galaxies \citep{Dressler_1999} in local galaxy clusters. 

Studies of AGN as a function of galaxy density and redshift are important as they give insights into the fuelling mechanisms behind AGN triggering. Models and simulations of galaxy formation currently require AGN feedback as an important mechanism for quenching star formation \citep[e.g.,][]{Croton_2006}, but the connection between AGN activity and large-scale galaxy environment is not fully understood. Recent studies show that clusters at high redshift appear to host more star formation and AGN activity compared to the local Universe \citep[e.g.,][]{Martini_2009, Galametz_2009, Bufanda_2016, Alberts_2016}. In addition, X-ray selected AGN are strongly clustered at $z\sim1$ \citep[e.g.,][]{Miyaji_2007, Bradshaw_2011}, and radio loud AGN (RLAGN) preferentially reside in denser environments at high redshift, compared to similarly massive non-active galaxies \citep{Hatch_2014}. Previous studies have found an increasing AGN fraction in clusters with redshift up to \mbox{$z\sim1.25$} \citep{Martini_2013, Kocevski_2009}. However, studies at \mbox{$z>1.5$} have been limited to investigating X-ray emission from protocluster galaxies selected based on techniques using rest-frame UV light, such as the BX/MD colour-colour methods \citep[see e.g.,][]{Adelberger_2004,Steidel_2003,Steidel_2004}, Lyman-alpha emitters (LAEs), and Lyman-break galaxies \citep[LBGs,][]{Lehmer_2009, Digby-North_2010, Lehmer_2013, Saez_2015}. This means that only limited (star-forming) protocluster galaxies were investigated, potentially biasing the AGN fraction if there is a strong dependence of AGN activity on host galaxy type. In addition, most of these studies cannot readily be compared to cluster AGN fractions at lower redshifts, as the X-ray observations are not deep enough to match the lower luminosity cuts in lower redshift studies.

In this paper, we present a comparison of the AGN fractions and AGN properties in the \clustername\ protocluster at $z=1.6233$, and a control field sample. The \clustername\ protocluster \citep{Papovich_2010,Tanaka_2010} is an ideal high-redshift structure to probe AGN activity due to the deep multiwavelength data available. This protocluster benefits from 14 band photometry and a clean yet complete sample of protocluster members \citep{Hatch_2016}, as well as sensitive \chandra\ data allowing us to probe X-ray luminosities as faint as $10^{43}$ erg s$^{-1}$ at $z\sim1.6$. 

The outline of the paper is as follows. We describe the data in Section~\ref{sec:data}. In Section~\ref{sec:AGN_activity}, we calculate AGN fractions and spatial distributions using uniformly selected X-ray AGN in cluster and field samples. In Section~\ref{sec:comparison_properties}, we compare the properties of protocluster AGN and field AGN. A discussion of the evolution of the AGN fraction in (proto)clusters from $z\sim3.09$ to $z\sim0.25$ follows in Section~\ref{sec:discussion}. We adopt a WMAP9 cosmology \citep{Hinshaw_2013}, with $\Omega_{\rm{m}} = 0.29$, $\Omega_\Lambda = 0.71$, and $h = 0.69$. All magnitudes are in the AB system. All X-ray luminosities quoted are calculated in rest-frame bands using a power-law model with a photon index $\Gamma=1.7$ to be consistent with comparison work \citep{Martini_2013}. We note that the effect of Galactic absorption on our fluxes is negligible.

\section{Description of the data}
\label{sec:data}

\subsection{X-ray selected AGN}
\label{sec:xray_sample}

We have selected our AGN using X-ray point source matching and a full band ($0.5$--$7$\,keV) X-ray luminosity cut of $L_X>10^{42}$ erg s$^{-1}$. We make use of \chandra\ X-ray imaging from the X-UDS program (PI: G.~Hasinger; Kocevski et al. in prep), which covers the central $0.33$ deg$^2$ of the UKIDSS Ultra Deep Survey (UDS) field (Almaini et al. in prep; described in Section~\ref{sec:control_field_sample}). The coverage includes the section of the UDS field that was observed as part of the Cosmic Assembly Near-infrared Deep Extragalactic Legacy Survey (CANDELS) with the \hubble\ \citep{Grogin_2011, Koekemoer_2011}. The X-UDS survey consists of $25$ ACIS-I pointings with a total integration time of 1.25 Ms.  The observations are tiled in a mosaic to achieve an average depth of $\sim600$ ksec in the central CANDELS region and $\sim200$ ksec in the remainder of the field.  The most recent X-ray point source catalogue contains $868$ unique detections. A threshold was applied to avoid false point source detections and to select only sources detected in any band with a false detection probability less than $1\times10^{-4}$, corresponding to $3.7\sigma$ detections and above.  Further details are provided in Kocevski et al. (in prep). This deep catalogue enables us to identify X-ray selected AGN at faint X-ray luminosities ($L_X \lesssim 10^{44}$~erg~s$^{-1} $), at the redshift of the protocluster. 

Using the maximum likelihood algorithm described in \citet{Civano_2012}, an optical counterparts catalogue was created for the CANDELS region of the X-UDS field. In this work we assume that optical/infrared sources within 1 arcsec of X-ray point sources are AGN for both cluster and control field samples (described in Section~\ref{sec:cluster_sample} and Section~\ref{sec:control_field_sample} respectively). We compare the AGN to the optical counterparts catalogue and find that this method is robust; $6/6$ protocluster AGN and $20/20$ field AGN within the CANDELS region are identical to the counterparts catalogue.

We adopt a full band ($0.5$--$7$\,keV) X-ray flux limit of \mbox{$6\times10^{-16}$~erg/s/cm$^2$}, defined using the flux limit map for the corresponding band. We choose this conservative value because the protocluster lies towards the edge of the \chandra\ field, so we take care to ensure a uniform flux limit for the control field and protocluster region. Figure~\ref{fig:clus_view} shows that the protocluster lies in a region of varying flux limit, and we find that the flux limit of \mbox{$6\times10^{-16}$~erg/s/cm$^{2}$} maximises both the depth of the data, and the coverage area available for the both cluster and control field samples. We test all the results presented in this paper using various flux limits and find that the results are consistent within quoted uncertainties.

\subsection{Cluster sample} 
\label{sec:cluster_sample}

\begin{figure*}
    \centering
    \begin{minipage}{\textwidth}
	\centering
	\includegraphics[width=\textwidth,trim=0.0cm 0cm 0.0cm 0.0cm, clip=true]{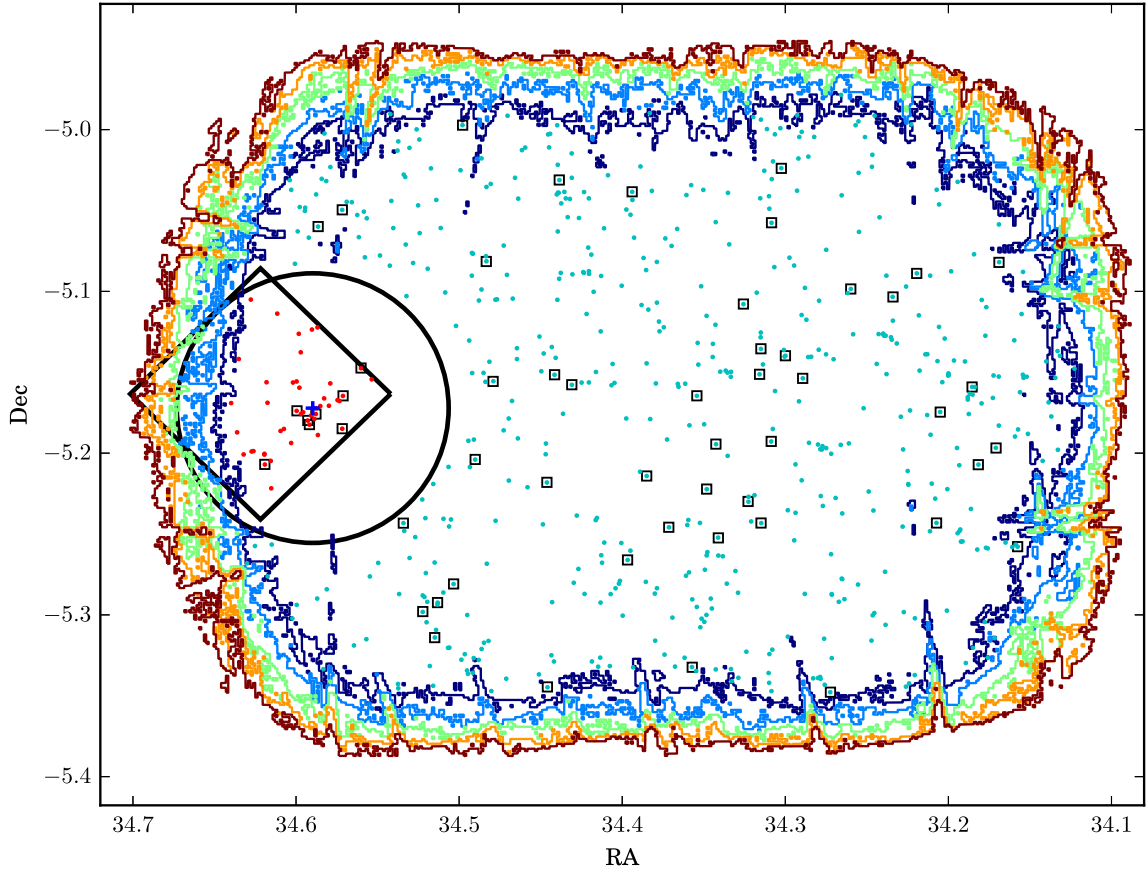}
\end{minipage}
\vspace*{-0.6cm}
\caption{This figure shows the distribution of cluster and field samples in the sky. The ``Goldilocks'' protocluster sample with $M_*>10^{10}$\,M$_{\odot}$ is shown as red points, protocluster AGN  as red points with black squares, control field galaxies as cyan points, and control field AGN as cyan points with black squares. The plotted galaxies are only those within the adopted flux limit, and the navy, light blue, green, orange, and red contour lines depict flux limits of $(6, 7, 8, 9, $ and $10) \times10^{-16}$ erg/s/cm$^{2}$ respectively. The diamond represents the field of view of the narrowband images of the protocluster, and the BCG is marked by the blue cross. The circle represents the excluded region in order to avoid contaminating the field sample with protocluster galaxies, and the empty white space within the circle (but not within the diamond), is a region where protocluster membership is unclear because it lacks the narrowband data.} 
\label{fig:clus_view}
\end{figure*}

For our cluster sample, we use the \clustername\ protocluster at $z=1.6233$ in the UDS field \citep{Papovich_2010, Tanaka_2010}. \citet{Hatch_2016} obtain multiwavelength imaging for this protocluster, including doubly sampled \textit{J} and \textit{K} imaging, in addition to imaging in two narrow-band filters (ESO/VLT FORS [SIII]+65 and HAWK-I $1.06\mu$m (NB1.06)). The field of view of the narrowband images is marked by the diamond in Figure~\ref{fig:clus_view}. The two narrowband filters were chosen such that they bracketed the Balmer break and the $4000$\,{\AA} break of the protocluster galaxies. This enabled the calculation of accurate photometric redshifts and stellar masses. \citet{Hatch_2016} found that, for 16 protocluster members with existing spectroscopic redshifts, the dispersion of $z_{\rm{phot}}-z_{\rm{spec}}$ was $\Delta z$/$(1+z) = 0.013$. This high-precision redshift data enables the accurate selection of protocluster members using photometry. The ``Goldilocks'' sample from \citet{Hatch_2016} consists of protocluster member galaxies that have been optimised to minimise contamination from field galaxies, as well as maximise completeness. Protocluster members in this sample were defined out to 5 arcmin (2.6 physical Mpc) apertures. Beyond this radius, the probability of Goldilocks protocluster candidates becoming cluster members is below 50\% and decreases rapidly with radius \citep{Hatch_2016}. 

Protocluster galaxy properties such as redshifts and masses have been determined through SED-fitting using \citet{Bruzual_Charlot_2003} stellar population templates and assuming a \citet{Chabrier_2003} initial mass function (IMF) as described in \citet{Hatch_2016}. The comoving volume of the full protocluster is $10.2\times10.2\times34.0$ Mpc$^3$ \citep{Hatch_2016}. Adopting the flux limit defined in Section~\ref{sec:xray_sample} results in a comoving volume of $2600$ Mpc$^3$ for the protocluster sample. 

To define our cluster sample, we use only the most massive members ($M_*>10^{10}$\,M$_{\odot}$) of the Goldilocks sample. We account for two effects before defining this mass cut. Firstly, we calculate the limiting mass of a galaxy starting from the X-ray flux, assuming that accretion on to the SMBH is at the Eddington rate, and that the bolometric correction from X-ray to total light is a factor of $10$. We find that this is $\sim10^9$M$_{\odot}$, assuming that the mass of the black hole is equal to $0.15\%$ of the mass of the host galaxy \citep{Kormendy_2000}. Secondly, the $99\%$ flux completeness limit for red galaxies at $z=1.62$ is $M_*>10^{9.7}$\,M$_{\odot}$ \citep{Hatch_2017}. 

The final cluster sample contains $46$ massive protocluster galaxies, which are shown as red points in Figure~\ref{fig:clus_view}, where AGN are highlighted with black squares. We find $8$ X-ray selected AGN in the Goldilocks protocluster sample, out of which 6 AGN have secure spectroscopic redshifts. We note that a further possible AGN is located at $\rm{RA} = 2{\rm{h}}\ 18{\rm{m}}\ 21.0{\rm{s}}$, $\rm{Dec} = -5^{\circ}\, 10'\, 20.1''\,$, the detection of which depends sensitively on the source detection parameters. We discard this source from our AGN sample, however, as this object was not detected a-priori in our \chandra\ source catalogue. There is also a further AGN at $\rm{RA} = 2{\rm{h}}\ 18{\rm{m}}\ 21.8{\rm{s}}$, $\rm{Dec} = -5^{\circ}\, 14'\, 55.3''\,$, that has been spectroscopically identified as a protocluster member (Hasinger et al. in prep). As this galaxy is outside the narrowband field of view, it is not part of the Goldilocks sample and, as a consequence, this AGN is not included in our analysis.

\subsection{Control field sample}
\label{sec:control_field_sample}

The data used are from the 8th data release of the Ultra Deep Survey (UDS DR8), which is a deep, photometric survey over an area of 0.8 sq degrees. The near-infrared data are some of the deepest over such a large area, reaching AB magnitude depths of $J = 24.9$, $H = 24.2$ and $K = 24.6$ \citep[see e.g.,][]{Hartley_2013,Simpson_2013}. We ensure that the field sample is not affected by the presence of the protocluster by excluding a circular region within 5 arcmin (corresponding to 6.80 comoving Mpc) of the protocluster centre. The Brightest Cluster Galaxy (BCG), marked by the blue cross, and the corresponding circle, are shown in Figure~\ref{fig:clus_view}. The BCG, with co-ordinates $\rm{RA} = 2{\rm{h}}\ 18{\rm{m}}\ 21.5{\rm{s}}$, $\rm{Dec} = -5^{\circ}\, 10'\, 19.8''\,$, is taken to be the highest mass galaxy in the protocluster member sample. 

To create a similarly selected field comparison sample, we select galaxies more massive than $10^{10}$\,M$_{\odot}$ with photometric redshifts in the range $1.5 < z < 1.7$. Masses and photometric redshifts in the UDS DR8 catalogue have also been determined using SED fitting \citep{Simpson_2013}, also using \citet{Bruzual_Charlot_2003} stellar population templates and assuming a \citet{Chabrier_2003} IMF. The selected field sample is as complete as the ``Goldilocks'' protocluster sample, and contains $550$ galaxies that are shown in cyan points in Figure~\ref{fig:clus_view}, with the $46$ AGN depicted as black squares. The size of the field region ($0.146$ deg$^2$) was kept as high as possible to maximise the sample size, while ensuring uniform X-ray coverage. The comoving volume of the field sample, taking into consideration the flux limit, is $\sim 350,000$ Mpc$^3$.

\section{AGN activity in clusters and field}
\label{sec:AGN_activity}

\subsection{AGN overdensity in protocluster}
\label{sec:AGN_overdensity}
\begin{figure}
    \begin{minipage}{0.48\textwidth}
	\centering
	\includegraphics[width=\columnwidth,trim=0.4cm 0.0cm 1.2cm 0.0cm, clip=true]{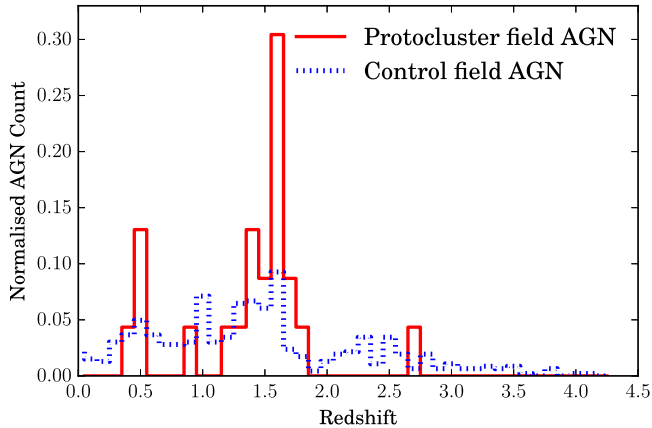}  
\end{minipage}
\vspace{-0.2cm}
\caption{Photometric redshift distribution of protocluster field AGN (in red) and control field AGN (in blue), normalised by the total number of AGN in their respective samples. At the redshift of the protocluster, $z=1.62$, there is a clear excess of AGN.} 
\label{fig:z_dist}
\end{figure}

We first investigate the abundance of AGN in the \clustername\ protocluster. Plotted in Figure~\ref{fig:z_dist} is the photometric redshift distribution of AGN (within the flux limit region) in the control field (blue dashed line) and in the narrowband field around the protocluster (red solid line), normalized by the total number of AGN in their respective samples. There is a clear excess of AGN at the redshift of the protocluster ($z \sim 1.62$), in the protocluster field compared to the control field, suggesting that there is indeed an overdensity of AGN associated with the protocluster. 

The AGN density in the protocluster is $(3.13 \pm 1.11) \times 10^{-3}$ Mpc$^{-3}$, and that of the field is $(1.33 \pm 0.20) \times 10^{-4}$ Mpc$^{-3}$. Errors are calculated using Poisson statistics. Thus the overdensity in the protocluster is $23\pm9$ times the field density. We perform a robustness check on this result as described in Section~\ref{sec:robustness}.

\begin{figure}
\includegraphics[width=\columnwidth,trim=0.5cm 0.0cm 1.2cm 0cm]{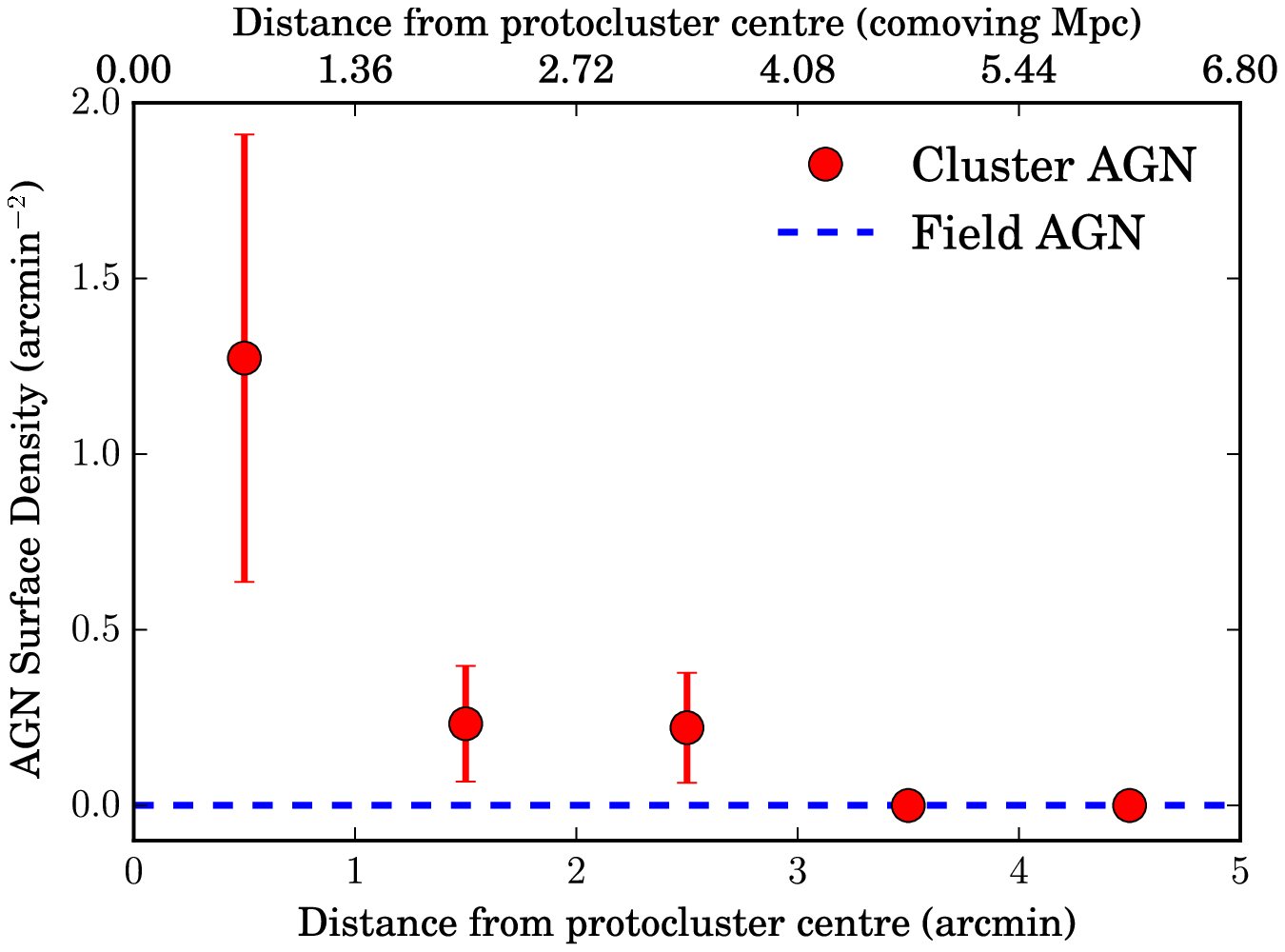}
\vspace*{-0.5cm}
\caption{Radial plot of AGN surface density in the field (blue dashed line), and protocluster (red circles). There is a significant surface density of AGN in the central arcmin of the protocluster. The field value has been normalised to account for the difference in comoving volumes between the control field and the protocluster.}
\label{fig:radial_plot}
\end{figure}

As seen in Figure~\ref{fig:clus_view}, the AGN are concentrated around the BCG (marked by the blue cross). We plot the AGN surface density as a function of distance from the BCG in Figure~\ref{fig:radial_plot}. The field value has been normalised to account for the difference in comoving volumes between the control field and the protocluster. We find that the AGN overdensity is present in the protocluster up until 3 arcmin, although it is most significant within the central arcmin of the protocluster. 

\subsection{Fraction of AGN in $M_*>10^{10}\rm\,M_{\odot}$ galaxies}
\label{sec:AGN_fraction}
We find an overdensity of AGN by a factor of $\sim23$ in the protocluster relative to the field. This overdensity could be because there is a higher AGN fraction in protoclusters, or simply because protoclusters contain a higher fraction of massive galaxies, which are more likely to host AGN \citep{Hatch_2011, Cooke_2014}. Therefore, we calculate, in both protocluster and field environments, the fraction of massive galaxies ($M_*>10^{10}$\,M$_{\odot}$) that are AGN. We find that the AGN fraction in the protocluster is $0.17^{+0.06}_{-0.05}$, while that of the field is $0.08 \pm 0.01$, meaning that the fraction of massive galaxies that host AGN in the protocluster is double that of the field. The errors are obtained using Wilson intervals, where the uncertainty $\delta{f_{i}}$ ($f_{i} = N_{i}/N_{\rm tot}$) is determined using the \citet{Wilson_1927} binomial confidence interval

\begin{equation}
\label{Equation: Frequency error}
f_{i} \pm \delta{f_{i}} = \frac {N_{i} + \kappa^2/{2}}{N_{\rm tot} + \kappa^2} \pm 
\frac{\kappa\sqrt{N_{\rm tot}}}{N_{\rm tot} + \kappa^2}\sqrt{f_{i}(1- f_{i}) + \frac{\kappa^2}{4N_{\rm tot}}},
\end{equation}

where $\kappa$ is the $100(1-\alpha/2)\rm{th}$ percentile of a standard normal distribution ($\alpha$ is the error percentile corresponding to the $1\sigma$ level; see \citealt{Brown_2001} for further details). We obtain an AGN enhancement in the protocluster at $1.6\sigma$ significance, and the errors are large as the sample size is small.

\begin{figure}
\includegraphics[width=\columnwidth,trim=0.6cm 0cm 1.2cm 0cm]{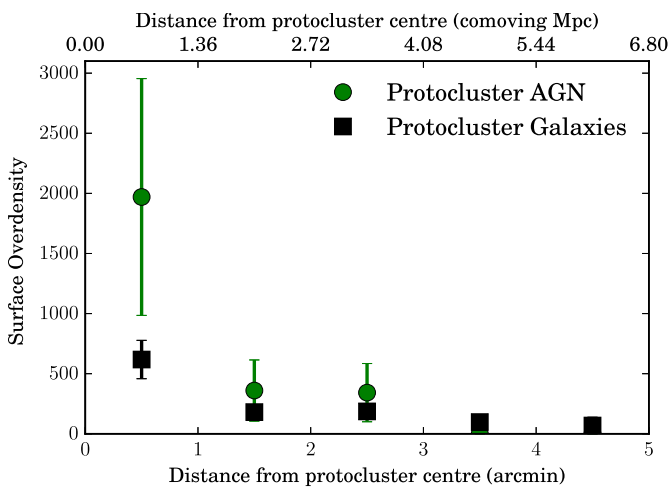}
\vspace*{-0.5cm}
\caption{Radial plot of the surface overdensity of protocluster AGN and protocluster galaxies. The green circles indicate the protocluster AGN surface density divided by the field AGN surface density, and the black squares indicate the protocluster galaxy surface density divided by the field galaxy surface density for massive galaxies ($M_*>10^{10}$\,M$_{\odot}$). There is a slight relative excess of AGN surface density compared to massive galaxy surface density, particularly in the central arcmin of the galaxy protocluster.}
\label{fig:norm_radial_plot}
\end{figure}

We also investigate whether the central concentration of AGN we find in Figure~\ref{fig:radial_plot} could be attributed to the distribution of massive galaxies within the protocluster. In Figure~\ref{fig:norm_radial_plot}, we plot the surface overdensity,

\begin{equation}
\label{Equation: Surface overdensity}
\rm{Surface\ Overdensity} = \frac {\rm{Protocluster\ Surface\ Density}}{\rm{Field\ Surface\ Density}}
\end{equation}

\noindent of both AGN and galaxies in the protocluster, as a function of the radius from the BCG. The green circles show the density excess of protocluster AGN as a function of radius from the BCG, and the black squares show the density excess of massive protocluster galaxies ($M_*>10^{10}$\,M$_{\odot}$). This figure shows that there is indeed a higher number of massive galaxies in the core of the protocluster relative to the field, but there is a slightly greater enhancement in the AGN fraction. However, as the number statistics are low, a larger sample of clusters is required to test the significance of this result.
 
In conclusion, there is an enhancement of AGN activity in this protocluster by a factor of $2.1\pm0.7$, above and beyond the overdensity of massive galaxies. This enhancement lies within 3 arcmin and mainly within the central arcmin of the protocluster (1 arcmin corresponds to 1.36 comoving Mpc). 

\section{Comparison between properties of protocluster and field AGN} 
\label{sec:comparison_properties}
We compare the properties of protocluster AGN to field AGN to see if the excess of AGN we find in the protocluster is correlated with differences in their properties, and to investigate whether environment affects the properties of these AGN. We test the null hypothesis that the distributions of the properties of field and protocluster AGN are sampling the same underlying distributions using Kolmogorov-Smirnov (KS) tests.

\begin{figure}
\includegraphics[width=\columnwidth,trim=0.6cm 0cm 1.2cm 0cm]{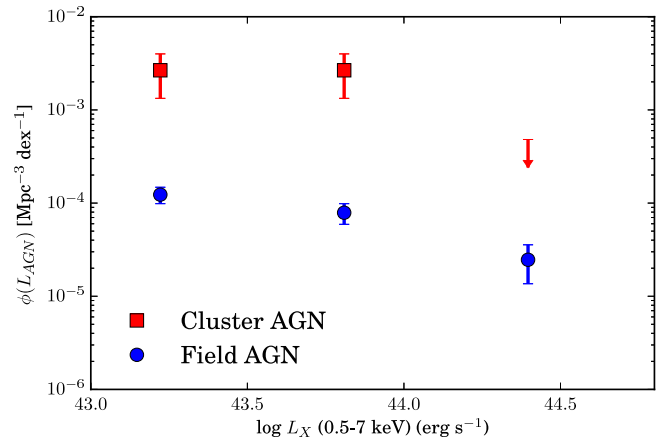}
\vspace*{-0.5cm}
\caption{Full band ($0.5$--$7$\,keV) X-ray luminosity function for AGN in the protocluster (red squares) and field (blue circles). Errors are calculated using Poisson statistics.}
\label{fig:lumfun}
\end{figure}

Firstly, X-ray luminosity functions were produced in order to compare the X-ray properties of field and protocluster AGN. X-ray luminosities were calculated using the X-ray fluxes in the full band ($0.5$--$7$\,keV). The luminosity functions, as shown in Figure~\ref{fig:lumfun}, were computed using the number of AGN corresponding to each luminosity bin within the comoving volume of the sample. Comparing AGN number densities at different X-ray luminosities allows us to compare the accretion rates in the two populations. The number density of protocluster AGN is, on average, $28 \pm 6$ times higher than that of the field AGN in the range of $10^{43}$ to $10^{45}$ erg s$^{-1}$, confirming the level of overdensity found in Section~\ref{sec:AGN_overdensity}. We observe that the X-ray luminosity functions of protocluster and field AGN appear to have the same shape. We test the null hypothesis that the individual X-ray luminosities are sampling the same underlying distribution using a KS test, resulting in $p = 0.82$. Therefore, we find that the shapes of the X-ray luminosity distributions are indistinguishable, and we find no evidence to suggest that the distributions of accretion rates of field and protocluster AGN are different.

Secondly, we examined the hardness ratio (HR), defined by,

\begin{equation}
 HR=\frac{h-s}{h+s},
\end{equation}

where $h$ is the flux in the hard band ($2$--$10$\,keV)and $s$ is the flux in the soft band ($0.5$--$2$\,keV). This was done in order to compare the obscuration by gas in field and protocluster AGN; more obscured AGN result in soft X-rays being absorbed. A KS test on the HR of the two populations does not show a significant difference; with $p = 0.22$. Therefore, this implies that the obscuration by gas within AGN does not significantly differ between AGN in field and protocluster environments.

Thirdly, we investigated the X-ray luminosity to stellar mass ratio of protocluster and field AGN. A lower ratio might imply that the AGN are running out of fuel, or that they are accreting less efficiently. The probability that the populations sample the same underlying distribution is $p=0.93$. Therefore, we find no evidence that protocluster and field AGN are at different stages of fuel consumption.

\begin{figure*}
    \centering
    \begin{minipage}{.48\textwidth}
        \centering
        \includegraphics[width=\columnwidth,trim={0.75cm 0 0.9cm 0},clip]{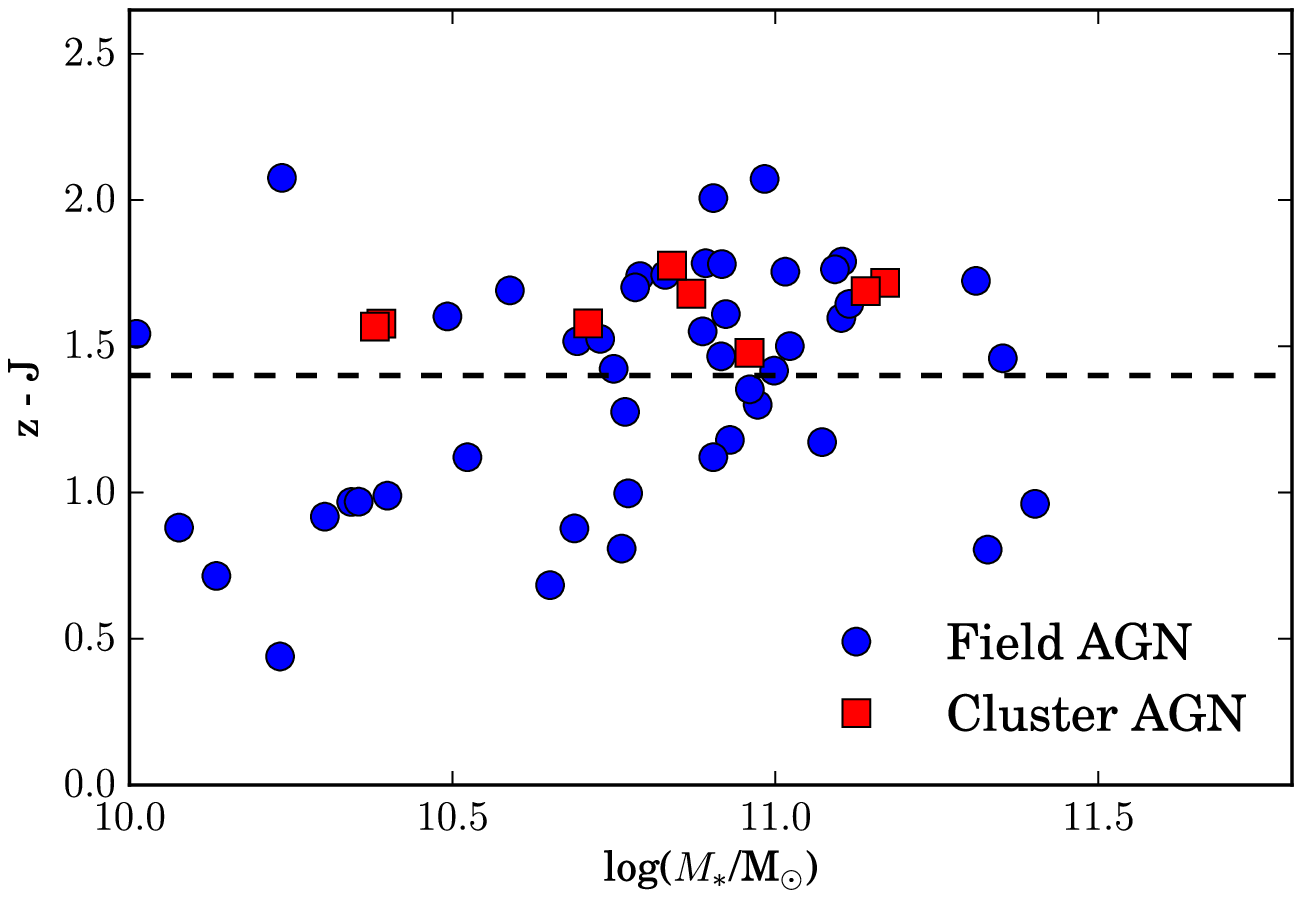}       
    \end{minipage}
    \begin{minipage}{.48\textwidth}
        \centering
        \includegraphics[width=\columnwidth,trim={0.25cm 0 1.4cm 0},clip]{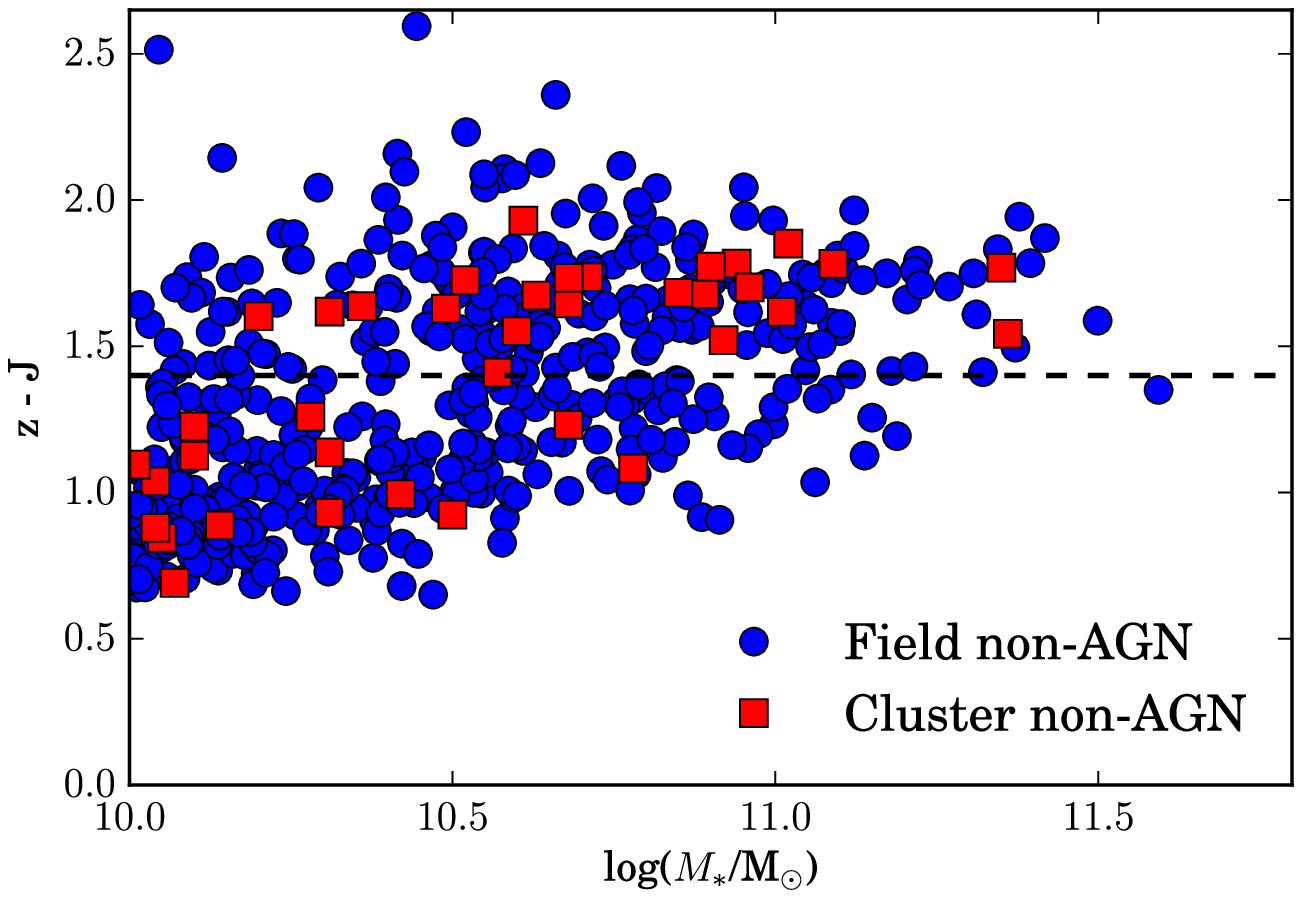}        
\end{minipage}
\vspace*{0cm}
\caption{\textbf{Left:} (a) $z-J$ colour--mass diagram for field AGN (blue circles) vs protocluster AGN (red squares). \textbf{Right:} (b) $z-J$ colour-mass diagram for field non-AGN (blue circles) vs protocluster non-AGN (red squares). The black dashed line shows $z-J=1.4$.}
\label{fig:colour_mass}
\end{figure*}

The observed $z-J$ colour corresponds to the rest-frame \mbox{$U-B$} colour, bracketing the $4000$\,{\AA} break \citep{Papovich_2010}. It is thus a proxy for mean stellar age of the galaxies, although it is also affected by dust obscuration. Using a colour cut of \mbox{$z-J > 1.4$} to define red galaxies, Figure~\ref{fig:colour_mass}(a) shows that the colours of protocluster AGN (red squares) are significantly redder than field AGN (blue circles). The probability that the colours of protocluster AGN and field AGN are drawn from the same distribution is $p = 0.03$ as given by a KS test. 

The protocluster galaxy population as a whole, however, is redder than the field, so the difference in colour between AGN in the protocluster and AGN in the field could be due to the environment and not the AGN. In Figure~\ref{fig:colour_mass}(b), we plot the colour--mass diagram of galaxies within the protocluster that do not host AGN and field galaxies that do not host AGN. We find that $100\%$ of our protocluster AGN are red, whereas only $57\%$ of protocluster non-AGN are red. However, a KS test on the $z-J$ colours of protocluster AGN and non-AGN results in $p=0.14$, and on field AGN and non-AGN results in $p=0.35$. Therefore, with the current data, we find no significant evidence that the $z-J$ colours differ from those of normal galaxies, in both environments. Hence, although the colours of AGN are redder in the protocluster as compared to the field, this is possibly due to the fact that protocluster galaxies are redder than field galaxies.

In conclusion, we find that the environment does not appear to impact most of the properties of AGN. We find no evidence that the properties of field and protocluster AGN differ significantly in terms of stellar mass distribution, hardness ratio, and X-ray luminosities. We find that colour is the only property affected by the different environments, as we find a significant difference between the colours of AGN in the protocluster and the field. However, we also find that the colours of field and protocluster AGN are not significantly different from typical field and protocluster galaxies, so these properties appear to randomly sample their parent distributions. In summary, by comparing the properties between field and protocluster AGN, we find no significant evidence that they are different. As there is no compelling theoretical reason to assume that the processes responsible for triggering/fuelling AGN activity are different in these two environments, we suggest that these processes simply occur more frequently in dense environments. Our study is based on a small AGN sample within a single protocluster, however, so larger sample sizes will be required to verify these interpretations.

\section{Robustness}
\label{sec:robustness}
It is important to consider that our analyses are affected by the issue of completeness of protocluster membership. This is a result of the selection technique used to define the ``Goldilocks'' sample, which depends on both galaxy magnitude and colour. Redder galaxies are fainter than bluer galaxies of the same mass, resulting in broader redshift probability functions due to higher fractional flux errors. Hence, fainter and redder galaxies are less likely to be selected as protocluster galaxies \citep{Hatch_2017}. As described in Section~\ref{sec:comparison_properties}, we found that protocluster AGN are redder than field AGN, with $100\%$ of protocluster AGN being red ($z-J > 1.4$), and the probability of sampling the same underlying colour distributions being $p=0.03$ as given by a KS test. Although these red AGN host galaxies are less likely to be selected as part of the ``Goldilocks'' sample, we still see an excess of protocluster AGN over the field AGN. 

However, we perform a robustness check, disregarding the ``Goldilocks'' sample, to test the AGN enhancement in the protocluster. We define the test protocluster sample to be massive galaxies ($M_*>10^{10}$\,M$_{\odot}$) in the UDS field with redshifts at $1.5<z<1.7$, within a circle of radius 5 arcmin centred on the BCG. We find $14$ AGN in this protocluster region (within the flux limit area), and subtract off the number of AGN in the field corresponding to the same area. We find a formal excess of $8.04$ AGN, and assume that this is associated with the protocluster. This is consistent with the $8$ AGN found using the ``Goldilocks'' sample in Section~\ref{sec:cluster_sample}. Therefore, we conclude that our result of the AGN enhancement in the protocluster is robust to the protocluster member selection technique used in \citet{Hatch_2017}. 

We find an AGN fraction of $0.17^{+0.06}_{-0.05}$ in massive protocluster galaxies, as described in Section~\ref{sec:AGN_fraction}. This fraction is likely to be robust as there are no significant differences in the $z-J$ colours of protocluster galaxies and protocluster AGN, as found in Section~\ref{sec:comparison_properties}. 

We also find that the AGN picked out as part of the test protocluster sample, and not the ``Goldilocks'' sample, are either massive and red, or blue. They all lie within the region of the colour-mass diagram where $>75\%$ of protocluster members would be correctly identified \citep{Hatch_2017}. It is therefore unlikely that we are missing any protocluster AGN due to the protocluster membership selection criterion. 

\section{Discussion}
\label{sec:discussion}
In summary of our results, we find that the AGN fraction in the $z\sim1.62$ protocluster is twice that of the field, and that the AGN enhancement lies within the central 3 arcmin (4.08 comoving Mpc) region of the protocluster. We find that the properties of field and protocluster AGN are not significantly different. As there is no significant evidence suggesting that they are triggered/fuelled in different ways, we infer that the processes responsible for triggering/fuelling AGN are possibly more frequent in denser environments.

\begin{figure}
\includegraphics[width=\columnwidth,trim=0.6cm 0cm 1.2cm 0cm]{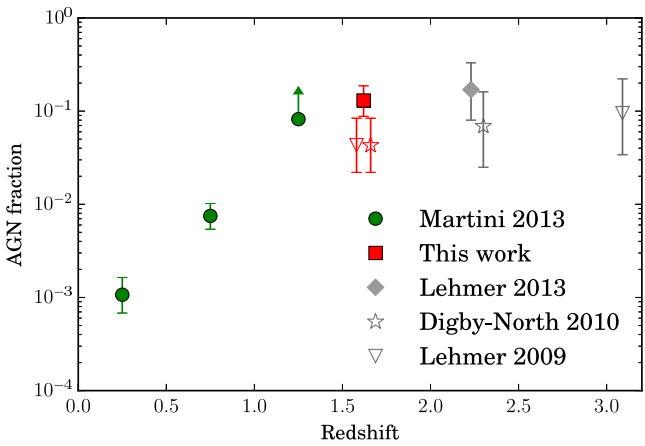}
\vspace*{-0.5cm}
\caption{Cluster AGN fraction ($L_X > 10^{43}$ erg s$^{-1}$) as a function of redshift. Data points from literature at $z<1.5$ are represented as green circles, and our work is represented as the red filled square. We plot the AGN fraction ($0.130^{+0.057}_{-0.042}$) using the hard band X-ray luminosity ($2$--$10$\,keV) here to be consistent with other works. The data at redshifts $0.25, 0.75, 1.25, 2.23, 2.30$ and $3.09$ are from \citet{Martini_2013}, \citet{Martini_2009}, \citet[][total AGN sample]{Martini_2013}, \citet[][HAE AGN sample]{Lehmer_2013}, \citet[][BX/MD AGN sample]{Digby-North_2010}, and \citet[][LBGs AGN sample]{Lehmer_2009} respectively. The three higher redshift studies are in grey as they do not sample the full protocluster galaxy population. The two highest redshift points are in open symbols as they use different luminosity cuts. We also calculate the AGN fraction at $z\sim1.62$ according to the luminosity limits used by the two higher redshift studies and plot them as red points with symbols corresponding to the studies. These have been offset slightly in redshift for clarity.}
\label{fig:clus_AGN_z}
\end{figure}

\begin{figure}
\includegraphics[width=\columnwidth,trim=0.6cm 0cm 1.2cm 0cm]{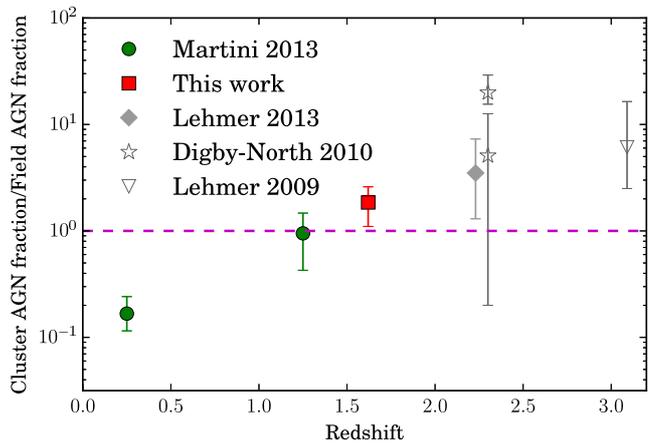}
\vspace*{-0.5cm}
\caption{Cluster AGN fraction relative to field AGN fraction as a function of redshift. There is a reversal in the local anti-correlation after $z>1.25$. The magenta dashed line indicates an equal cluster and field AGN fraction. As with Figure~\ref{fig:clus_AGN_z}, data points from literature at $z<1.5$ are represented as green circles, and this work (using the hard band X-ray luminosity) is represented as the red filled square. The three higher redshift studies are in grey as they do not sample the full protocluster galaxy population; \citet[][HAE AGN sample]{Lehmer_2013}, \citet[][higher and lower points are BX/MD AGN and emission line AGN respectively]{Digby-North_2010} and \citet[][mean AGN fraction among LBGs and LAEs]{Lehmer_2009}. Open symbols denote that the luminosity cuts are different to \mbox{$L_X > 10^{43}$~erg~s$^{-1}$}.}
\label{fig:relative_clus_AGN_z}
\end{figure}

To frame our results in the context of recent literature on (proto)clusters at higher and lower redshifts, we plot the AGN fraction and the ratio of cluster AGN fraction to field AGN fraction as a function of redshift in Figure~\ref{fig:clus_AGN_z} and Figure~\ref{fig:relative_clus_AGN_z} respectively. The cut in X-ray luminosity is $10^{43}$ erg s$^{-1}$, except for the two highest redshift studies at $z=2.30$ and $z=3.09$, in which the cuts are \mbox{$4.6\times10^{43}$~erg~s$^{-1}$} and $3.2\times10^{43}$ erg s$^{-1}$ respectively. Figure~\ref{fig:clus_AGN_z} shows that there is an increasing cluster AGN fraction with redshift. It rises to $\sim17\%$ at $z\sim1.6$ and then flattens; it is uncertain beyond $z\sim2$, however, because of the different luminosity limits applied. This increase in the cluster AGN fraction with redshift has also been found by several recent studies \citep[e.g.,][]{Galametz_2009,Martini_2009,Bufanda_2016,Alberts_2016}. The AGN fraction in the field, however, also increases with redshift \citep[e.g.,][]{Merloni_2013}. To study the influence of environment we compare the cluster AGN fractions to field AGN fractions. Figure~\ref{fig:relative_clus_AGN_z} shows that the relative AGN activity in clusters compared to the field increases with redshift. The AGN fraction in clusters is lower than the field at $z<1$, but we find a larger AGN fraction in the $z\sim1.62$ protocluster compared to the field. We therefore find evidence for a reversal in the local anti-correlation between galaxy density and AGN fraction, confirming the results of \citet{Martini_2013}.    

The AGN fraction in the $z\sim1.62$ protocluster is slightly higher than those of two protoclusters at $z=2.30$ \citep{Digby-North_2010} and $z=3.09$ \citep{Lehmer_2009}, even though it is consistent within error-bars. This is possibly because they use different techniques to select the protocluster members, such as Lyman-alpha emitters (LAEs), Lyman-break galaxies (LBGs), and BX/MD. These techniques will result in incomplete protocluster membership as they are biased towards strongly star-forming galaxies, and are likely to miss quiescent galaxies. In addition, their cuts in X-ray luminosities are higher than ours, possibly contributing to the (marginally) lower AGN fraction. We adopt their cuts in X ray luminosity, recalculate the protocluster AGN fraction, and plot these in Figure~\ref{fig:clus_AGN_z}. Figure~\ref{fig:relative_clus_AGN_z} shows that despite the lower cluster AGN fraction in these two protoclusters, the relative enhancement of cluster AGN over field AGN still increases with redshift. This may be because the luminosity cuts and the methods used for identifying galaxies are the same in the cluster and the field within each sample, and so studying the relative enhancment may be more appropriate for comparison between different studies. 

The AGN fraction in clusters at $z>1.5$ is $10$--$20\%$. This could mean that each massive protocluster galaxy is frequently ``switched on'' in terms of AGN activity, or that the phenomenon happens once but lasts for a longer time in the protocluster compared to the field. In Section~\ref{sec:comparison_properties}, we found that there are no significant differences in the properties of AGN between the two different environments and interpreted that there is no evidence suggesting that the mechanisms responsible for triggering/fuelling AGN are different in the protocluster compared to the field. Therefore, the mechanisms responsible may simply be more frequent in the protocluster environment than the field. 

Mergers and interactions such as galaxy harassment \citep{Moore_1996} have been suggested as the mechanisms responsible for triggering AGN activity \citep{Springel_2005}. These processes may provide the instabilities required to funnel gas towards the SMBH. The decrease in the overall AGN fraction over cosmic time could be due to a decrease in frequency of fuelling mechanisms or due to a decrease in the amount of fuel available. It has been found that the frequency of mergers involving massive galaxies ($M_*>10^{10}$\,M$_{\odot}$) decreases as the Universe ages \citep{Conselice_2007}. The cold gas supply is also depleted as the Universe ages as it forms stars and accretes on to black holes. The suppression of AGN activity in mature clusters relative to the field in the local Universe may be due to virialization, as this has been suggested to halt merger rates \citep{Lotz_2013}. It has been found by \citet{van_Breukelen_2009} that AGN are triggered by galaxy interaction and merging events during the pre-virialization evolutionary stage.

\citet{Lotz_2013} explore the frequency of mergers in the \clustername\ protocluster, and find that the merger rate for galaxies in the protocluster is $\sim2$--$4$ mergers per Gyr per galaxy, as compared to $\sim0.5$ mergers per Gyr per galaxy in the field. This increased merger rate may be responsible for the increase in AGN rates.

To test whether mergers and interactions are more frequent in the protocluster AGN compared to the field AGN, we used CANDELS-UDS visual classifications to calculate merger fractions using both the fraction of galaxies classed as ``irregular'', and those classed as ``disturbed'' (i.e. mergers or interactions). These morphologies were visually identified by a team of astronomers within the CANDELS collaboration \citep{Kartaltepe_2015}. We impose that $>50\%$ of classifiers must agree in order to accept the classification. We find that $4/6$ protocluster AGN are ``disturbed'', compared to $3/18$ field AGN ($67^{+16}_{-20}\%$ in the protocluster AGN as opposed to $17^{+10}_{-7}\%$ in the field AGN). The ``irregular'' fraction is $2/6$ in the protocluster AGN and $0/20$ in the field AGN ($33^{+20}_{-16}\%$ in the protocluster AGN and an upper limit of $5\%$ in the field AGN). Errors are calculated following \citet{Wilson_1927} as described in Section~\ref{sec:AGN_fraction}. We note that morphologies may be subjective, and thus conclude that there is tentative evidence that mergers and interactions are fuelling AGN in the $z\sim1.62$ protocluster. However, we also find that among the inactive galaxies in the protocluster, $18^{+14}_{-9}\%$ were classified as ``disturbed'', and an upper limit of $8\%$ were classified as ``irregular''. This provides more evidence to support the hypothesis that the enhancement in AGN correlates with mergers and environmental interactions. We also find that, among the ``disturbed'' galaxies, $4/6$ are AGN in the protocluster ($67^{+16}_{-20}\%$), while $3/28$ are AGN in the field ($11^{+7}_{-5}\%$). This may suggest that the protocluster environment enhances the probability that a merger/interaction triggers an AGN.   

We find that the AGN enhancement in the $z\sim1.62$ protocluster lies mainly in its central regions. An excess of AGN has also been found by \citet{Galametz_2009} in the central regions of clusters at lower redshifts. We find a larger excess in our study, however this is expected as the cluster AGN fraction increases with redshift as shown in Figure~\ref{fig:clus_AGN_z}. Star formation in clusters also increases with redshift, and this could point towards a co-evolution between star-formation activity and AGN activity. This may be expected because they share the same gas source that becomes depleted as the Universe ages. However, \citet{Hatch_2017} find that the central regions of the same protocluster at $z\sim1.62$ have suppressed sSFR compared to outer regions, and one of many possibilities is that AGN feedback quenches star formation.

The high AGN fraction in protoclusters at high redshift may have important implications for our understanding of galaxy evolution. A key ingredient in regulating star formation in current galaxy formation models is feedback from AGN \citep[e.g.,][]{Croton_2006}. Therefore, the high protocluster AGN fraction at $z\sim1.62$ could imply more rapid quenching of star formation in dense environments at high redshift. Yet in models of galaxy formation, no direct prescription for environmental dependence is applied to AGN feedback. Prescriptions of AGN feedback in some semi-analytic models do indirectly depend on environment, as clusters have larger halo masses so there is more gas mass available for fuelling AGN, and as there is an environmental dependence of mergers, which stimulate accretion onto SMBHs \citep{Henriques_2016}. However, environmental interactions such as harassment \citep{Moore_1996} could also disturb protocluster environments to funnel gas onto the SMBH (without a merger), and thus mass-quench galaxies in denser environments. It has been proposed that ``mass quenching'' (e.g. AGN feedback) and ``environmental quenching'' (e.g. mergers) are mechanisms that extinguish star formation independent of each other \citep[e.g.,][]{Peng_2010}. However, in this work we find evidence for environmental dependence of AGN activity, consistent with recent work \citep{Darvish_2016} that finds an environmental dependence on mass quenching efficiency. This may therefore be evidence that a more direct environmental dependence of AGN feedback must be applied in galaxy formation models, as quenching mechanisms are crucial in determining galaxy formation and evolution. 

\section{Summary}

In this work we study the prevalence of X-ray AGN in the \clustername\ protocluster at $z=1.6233$, and compare them to a control field sample at $1.5 < z < 1.7$. We investigate the properties of field and protocluster AGN, and study the evolution of AGN activity in dense environments over cosmic time. We confirm a reversal of the local anti-correlation between galaxy density and AGN activity, as suggested by \citet{Martini_2013}. To summarise our findings:

\begin{enumerate}

\item We find an overdensity of AGN in the protocluster relative to the field; $23\pm9$ times the number of AGN per unit volume. The AGN fraction of massive galaxies in the protocluster is $2.1\pm0.7$ times that of massive galaxies in the control field.

\item The AGN excess lies within 3 arcmin and mainly within the central arcmin of the protocluster. Therefore AGN activity is enhanced in the region of massive groups, where the sSFR of the galaxies is suppressed. 

\item We find that the properties of field and protocluster AGN are not significantly different in terms of stellar mass distribution, hardness ratio, and X-ray luminosity. In terms of colours and stellar masses, field and protocluster AGN are not significantly different to typical field and protocluster galaxies, respectively. We conclude that there is no evidence suggesting that AGN in different environments are triggered/fuelled in different ways, and infer that the processes that trigger/fuel AGN are simply more frequent in denser environments.

\item We use CANDELS visually classified morphologies to test whether environmental interactions could be triggering AGN. The morphologically classified disturbed and irregular fractions are higher in cluster AGN than field AGN. The more frequent mergers and environmental interactions in the protocluster could explain the enhancement of AGN activity.

\item  We combine our study with recent literature and find that the overall AGN fraction decreases with cosmic time. We find that the relative enhancement of cluster AGN and field AGN decreases as the Universe ages. 

\end{enumerate}

\section*{Acknowledgements}

We thank the anonymous referee for their insightful comments, which improved the paper. CK is supported by a Vice-Chancellor's Scholarship for Research Excellence. NAH acknowledges support from STFC through an Ernest Rutherford Fellowship. EAC acknowledges support from the ERC Advanced Investigator programme DUSTYGAL 321334. SIM acknowledges the support of the STFC consolidated grant ST/K001000/1 to the astrophysics group at the University of Leicester. This work is based on observations made with ESO Telescopes at the La Silla Paranal Observatory under programme ID 089.A-0126. This work also uses data from ESO telescopes at the Paranal Observatory (programmes 094.A-0410 and 180.A-0776; PI: Almaini). We are grateful to the staff at UKIRT for their tireless efforts in ensuring the success of the UDS project. We also wish to recognize and acknowledge the very significant cultural role and reverence that the summit of Mauna Kea has within the indigenous Hawaiian community. We were most fortunate to have the opportunity to conduct observations from this mountain.

\bibliographystyle{mnras}
\bibliography{biblio3}

\begin{thebibliography}{}
\makeatletter
\relax
\def\mn@urlcharsother{\let\do\@makeother \do\$\do\&\do\#\do\^\do\_\do\%\do\~}
\def\mn@doi{\begingroup\mn@urlcharsother \@ifnextchar [ {\mn@doi@}
  {\mn@doi@[]}}
\def\mn@doi@[#1]#2{\def\@tempa{#1}\ifx\@tempa\@empty \href
  {http://dx.doi.org/#2} {doi:#2}\else \href {http://dx.doi.org/#2} {#1}\fi
  \endgroup}
\def\mn@eprint#1#2{\mn@eprint@#1:#2::\@nil}
\def\mn@eprint@arXiv#1{\href {http://arxiv.org/abs/#1} {{\tt arXiv:#1}}}
\def\mn@eprint@dblp#1{\href {http://dblp.uni-trier.de/rec/bibtex/#1.xml}
  {dblp:#1}}
\def\mn@eprint@#1:#2:#3:#4\@nil{\def\@tempa {#1}\def\@tempb {#2}\def\@tempc
  {#3}\ifx \@tempc \@empty \let \@tempc \@tempb \let \@tempb \@tempa \fi \ifx
  \@tempb \@empty \def\@tempb {arXiv}\fi \@ifundefined
  {mn@eprint@\@tempb}{\@tempb:\@tempc}{\expandafter \expandafter \csname
  mn@eprint@\@tempb\endcsname \expandafter{\@tempc}}}

\bibitem[\protect\citeauthoryear{{Adelberger}, {Steidel}, {Shapley}, {Hunt},
  {Erb}, {Reddy}  \& {Pettini}}{{Adelberger} et~al.}{2004}]{Adelberger_2004}
{Adelberger} K.~L.,  {Steidel} C.~C.,  {Shapley} A.~E.,  {Hunt} M.~P.,  {Erb}
  D.~K.,  {Reddy} N.~A.,   {Pettini} M.,  2004, \mn@doi [\apj]
  {10.1086/383221}, 607, 226

\bibitem[\protect\citeauthoryear{{Alberts} et~al.,}{{Alberts}
  et~al.}{2016}]{Alberts_2016}
{Alberts} S.,  et~al., 2016, \mn@doi [\apj] {10.3847/0004-637X/825/1/72}, 825,
  72

\bibitem[\protect\citeauthoryear{{Boyle}, {Georgantopoulos}, {Blair},
  {Stewart}, {Griffiths}, {Shanks}, {Gunn}  \& {Almaini}}{{Boyle}
  et~al.}{1998}]{Boyle_1998}
{Boyle} B.~J.,  {Georgantopoulos} I.,  {Blair} A.~J.,  {Stewart} G.~C.,
  {Griffiths} R.~E.,  {Shanks} T.,  {Gunn} K.~F.,   {Almaini} O.,  1998,
  \mn@doi [\mnras] {10.1046/j.1365-8711.1998.01098.x}, 296, 1

\bibitem[\protect\citeauthoryear{{Bradshaw} et~al.,}{{Bradshaw}
  et~al.}{2011}]{Bradshaw_2011}
{Bradshaw} E.~J.,  et~al., 2011, \mn@doi [\mnras]
  {10.1111/j.1365-2966.2011.18888.x}, 415, 2626

\bibitem[\protect\citeauthoryear{Brown, Cai  \& DasGupta}{Brown
  et~al.}{2001}]{Brown_2001}
Brown L.~D.,  Cai T.~T.,   DasGupta A.,  2001, \mn@doi [Statist. Sci.]
  {10.1214/ss/1009213286}, 16, 101

\bibitem[\protect\citeauthoryear{{Bruzual} \& {Charlot}}{{Bruzual} \&
  {Charlot}}{2003}]{Bruzual_Charlot_2003}
{Bruzual} G.,  {Charlot} S.,  2003, \mn@doi [\mnras]
  {10.1046/j.1365-8711.2003.06897.x}, 344, 1000

\bibitem[\protect\citeauthoryear{{Bufanda} et~al.,}{{Bufanda}
  et~al.}{2017}]{Bufanda_2016}
{Bufanda} E.,  et~al., 2017, \mn@doi [\mnras] {10.1093/mnras/stw2824}, 465,
  2531

\bibitem[\protect\citeauthoryear{{Chabrier}}{{Chabrier}}{2003}]{Chabrier_2003}
{Chabrier} G.,  2003, \mn@doi [\pasp] {10.1086/376392}, 115, 763

\bibitem[\protect\citeauthoryear{{Civano} et~al.,}{{Civano}
  et~al.}{2012}]{Civano_2012}
{Civano} F.,  et~al., 2012, \mn@doi [\apjs] {10.1088/0067-0049/201/2/30}, 201,
  30

\bibitem[\protect\citeauthoryear{{Conselice}}{{Conselice}}{2007}]{Conselice_20%
07}
{Conselice} C.~J.,  2007, {Galaxy Mergers and Interactions at High Redshift}.
pp 381--384 (\mn@eprint {} {astro-ph/0610662}),
  \mn@doi{10.1017/S1743921306010222}

\bibitem[\protect\citeauthoryear{{Cooke}, {Hatch}, {Muldrew}, {Rigby}  \&
  {Kurk}}{{Cooke} et~al.}{2014}]{Cooke_2014}
{Cooke} E.~A.,  {Hatch} N.~A.,  {Muldrew} S.~I.,  {Rigby} E.~E.,   {Kurk}
  J.~D.,  2014, \mn@doi [\mnras] {10.1093/mnras/stu522}, 440, 3262

\bibitem[\protect\citeauthoryear{{Croton} et~al.,}{{Croton}
  et~al.}{2006}]{Croton_2006}
{Croton} D.~J.,  et~al., 2006, \mn@doi [\mnras]
  {10.1111/j.1365-2966.2005.09675.x}, 365, 11

\bibitem[\protect\citeauthoryear{{Darvish}, {Mobasher}, {Sobral}, {Rettura},
  {Scoville}, {Faisst}  \& {Capak}}{{Darvish} et~al.}{2016}]{Darvish_2016}
{Darvish} B.,  {Mobasher} B.,  {Sobral} D.,  {Rettura} A.,  {Scoville} N.,
  {Faisst} A.,   {Capak} P.,  2016, \mn@doi [\apj]
  {10.3847/0004-637X/825/2/113}, 825, 113

\bibitem[\protect\citeauthoryear{{Digby-North} et~al.,}{{Digby-North}
  et~al.}{2010}]{Digby-North_2010}
{Digby-North} J.~A.,  et~al., 2010, \mn@doi [\mnras]
  {10.1111/j.1365-2966.2010.16977.x}, 407, 846

\bibitem[\protect\citeauthoryear{{Dressler}, {Thompson}  \&
  {Shectman}}{{Dressler} et~al.}{1985}]{Dressler_1985}
{Dressler} A.,  {Thompson} I.~B.,   {Shectman} S.~A.,  1985, \mn@doi [\apj]
  {10.1086/162813}, 288, 481

\bibitem[\protect\citeauthoryear{{Dressler}, {Smail}, {Poggianti}, {Butcher},
  {Couch}, {Ellis}  \& {Oemler}}{{Dressler} et~al.}{1999}]{Dressler_1999}
{Dressler} A.,  {Smail} I.,  {Poggianti} B.~M.,  {Butcher} H.,  {Couch} W.~J.,
  {Ellis} R.~S.,   {Oemler} Jr. A.,  1999, \mn@doi [\apjs] {10.1086/313213},
  122, 51

\bibitem[\protect\citeauthoryear{{Farouki} \& {Shapiro}}{{Farouki} \&
  {Shapiro}}{1981}]{Farouki_Shapiro_1981}
{Farouki} R.,  {Shapiro} S.~L.,  1981, \mn@doi [\apj] {10.1086/158563}, 243, 32

\bibitem[\protect\citeauthoryear{{Ferrarese} \& {Merritt}}{{Ferrarese} \&
  {Merritt}}{2000}]{Ferrarese_2000}
{Ferrarese} L.,  {Merritt} D.,  2000, \mn@doi [\apjl] {10.1086/312838}, 539, L9

\bibitem[\protect\citeauthoryear{{Franceschini}, {Hasinger}, {Miyaji}  \&
  {Malquori}}{{Franceschini} et~al.}{1999}]{Franceschini_1999}
{Franceschini} A.,  {Hasinger} G.,  {Miyaji} T.,   {Malquori} D.,  1999,
  \mn@doi [\mnras] {10.1046/j.1365-8711.1999.03078.x}, 310, L5

\bibitem[\protect\citeauthoryear{{Galametz} et~al.,}{{Galametz}
  et~al.}{2009}]{Galametz_2009}
{Galametz} A.,  et~al., 2009, \mn@doi [\apj] {10.1088/0004-637X/694/2/1309},
  694, 1309

\bibitem[\protect\citeauthoryear{{Gebhardt} et~al.,}{{Gebhardt}
  et~al.}{2000}]{Gebhardt_2000}
{Gebhardt} K.,  et~al., 2000, \mn@doi [\apjl] {10.1086/312840}, 539, L13

\bibitem[\protect\citeauthoryear{{Giovanelli} \& {Haynes}}{{Giovanelli} \&
  {Haynes}}{1985}]{Giovanelli_1985}
{Giovanelli} R.,  {Haynes} M.~P.,  1985, \mn@doi [\apj] {10.1086/163170}, 292,
  404

\bibitem[\protect\citeauthoryear{{Gisler}}{{Gisler}}{1978}]{Gisler_1978}
{Gisler} G.~R.,  1978, \mn@doi [\mnras] {10.1093/mnras/183.4.633}, 183, 633

\bibitem[\protect\citeauthoryear{{Grogin} et~al.,}{{Grogin}
  et~al.}{2011}]{Grogin_2011}
{Grogin} N.~A.,  et~al., 2011, \mn@doi [\apjs] {10.1088/0067-0049/197/2/35},
  197, 35

\bibitem[\protect\citeauthoryear{{Gunn} \& {Gott}}{{Gunn} \&
  {Gott}}{1972}]{Gunn_Gott_1972}
{Gunn} J.~E.,  {Gott} III J.~R.,  1972, \mn@doi [\apj] {10.1086/151605}, 176, 1

\bibitem[\protect\citeauthoryear{{Hartley} et~al.,}{{Hartley}
  et~al.}{2013}]{Hartley_2013}
{Hartley} W.~G.,  et~al., 2013, \mn@doi [\mnras] {10.1093/mnras/stt383}, 431,
  3045

\bibitem[\protect\citeauthoryear{{Hatch}, {Kurk}, {Pentericci}, {Venemans},
  {Kuiper}, {Miley}  \& {R{\"o}ttgering}}{{Hatch} et~al.}{2011}]{Hatch_2011}
{Hatch} N.~A.,  {Kurk} J.~D.,  {Pentericci} L.,  {Venemans} B.~P.,  {Kuiper}
  E.,  {Miley} G.~K.,   {R{\"o}ttgering} H.~J.~A.,  2011, \mn@doi [\mnras]
  {10.1111/j.1365-2966.2011.18735.x}, 415, 2993

\bibitem[\protect\citeauthoryear{{Hatch} et~al.,}{{Hatch}
  et~al.}{2014}]{Hatch_2014}
{Hatch} N.~A.,  et~al., 2014, \mn@doi [\mnras] {10.1093/mnras/stu1725}, 445,
  280

\bibitem[\protect\citeauthoryear{{Hatch}, {Muldrew}, {Cooke}, {Hartley},
  {Almaini}, {Simpson}  \& {Conselice}}{{Hatch} et~al.}{2016}]{Hatch_2016}
{Hatch} N.~A.,  {Muldrew} S.~I.,  {Cooke} E.~A.,  {Hartley} W.~G.,  {Almaini}
  O.,  {Simpson} C.~J.,   {Conselice} C.~J.,  2016, \mn@doi [\mnras]
  {10.1093/mnras/stw602}, 459, 387

\bibitem[\protect\citeauthoryear{{Hatch}, {Cooke}, {Muldrew}, {Hartley},
  {Almaini}, {Conselice}  \& {Simpson}}{{Hatch} et~al.}{2017}]{Hatch_2017}
{Hatch} N.~A.,  {Cooke} E.~A.,  {Muldrew} S.~I.,  {Hartley} W.~G.,  {Almaini}
  O.,  {Conselice} C.~J.,   {Simpson} C.~J.,  2017, \mn@doi [\mnras]
  {10.1093/mnras/stw2359}, 464, 876

\bibitem[\protect\citeauthoryear{{Henriques}, {White}, {Thomas}, {Angulo},
  {Guo}, {Lemson}  \& {Wang}}{{Henriques} et~al.}{2016}]{Henriques_2016}
{Henriques} B.~M.~B.,  {White} S.~D.~M.,  {Thomas} P.~A.,  {Angulo} R.~E.,
  {Guo} Q.,  {Lemson} G.,   {Wang} W.,  2016, preprint (\mn@eprint {arXiv}
  {1611.02286})

\bibitem[\protect\citeauthoryear{{Hinshaw} et~al.,}{{Hinshaw}
  et~al.}{2013}]{Hinshaw_2013}
{Hinshaw} G.,  et~al., 2013, \mn@doi [\apjs] {10.1088/0067-0049/208/2/19}, 208,
  19

\bibitem[\protect\citeauthoryear{{Kartaltepe} et~al.,}{{Kartaltepe}
  et~al.}{2015}]{Kartaltepe_2015}
{Kartaltepe} J.~S.,  et~al., 2015, \mn@doi [\apjs]
  {10.1088/0067-0049/221/1/11}, 221, 11

\bibitem[\protect\citeauthoryear{{Kauffmann}, {White}, {Heckman}, {M{\'e}nard},
  {Brinchmann}, {Charlot}, {Tremonti}  \& {Brinkmann}}{{Kauffmann}
  et~al.}{2004}]{Kauffmann_2004}
{Kauffmann} G.,  {White} S.~D.~M.,  {Heckman} T.~M.,  {M{\'e}nard} B.,
  {Brinchmann} J.,  {Charlot} S.,  {Tremonti} C.,   {Brinkmann} J.,  2004,
  \mn@doi [\mnras] {10.1111/j.1365-2966.2004.08117.x}, 353, 713

\bibitem[\protect\citeauthoryear{{Kocevski}, {Lubin}, {Gal}, {Lemaux},
  {Fassnacht}  \& {Squires}}{{Kocevski} et~al.}{2009}]{Kocevski_2009}
{Kocevski} D.~D.,  {Lubin} L.~M.,  {Gal} R.,  {Lemaux} B.~C.,  {Fassnacht}
  C.~D.,   {Squires} G.~K.,  2009, \mn@doi [\apj]
  {10.1088/0004-637X/690/1/295}, 690, 295

\bibitem[\protect\citeauthoryear{{Koekemoer} et~al.,}{{Koekemoer}
  et~al.}{2011}]{Koekemoer_2011}
{Koekemoer} A.~M.,  et~al., 2011, \mn@doi [\apjs] {10.1088/0067-0049/197/2/36},
  197, 36

\bibitem[\protect\citeauthoryear{{Kormendy}}{{Kormendy}}{2000}]{Kormendy_2000}
{Kormendy} J.,  2000, ArXiv Astrophysics e-prints

\bibitem[\protect\citeauthoryear{{Larson}, {Tinsley}  \& {Caldwell}}{{Larson}
  et~al.}{1980}]{Larson_1980}
{Larson} R.~B.,  {Tinsley} B.~M.,   {Caldwell} C.~N.,  1980, \mn@doi [\apj]
  {10.1086/157917}, 237, 692

\bibitem[\protect\citeauthoryear{{Lehmer} et~al.,}{{Lehmer}
  et~al.}{2009}]{Lehmer_2009}
{Lehmer} B.~D.,  et~al., 2009, \mn@doi [\apj] {10.1088/0004-637X/691/1/687},
  691, 687

\bibitem[\protect\citeauthoryear{{Lehmer} et~al.,}{{Lehmer}
  et~al.}{2013}]{Lehmer_2013}
{Lehmer} B.~D.,  et~al., 2013, \mn@doi [\apj] {10.1088/0004-637X/765/2/87},
  765, 87

\bibitem[\protect\citeauthoryear{{Lotz} et~al.,}{{Lotz}
  et~al.}{2013}]{Lotz_2013}
{Lotz} J.~M.,  et~al., 2013, \mn@doi [\apj] {10.1088/0004-637X/773/2/154}, 773,
  154

\bibitem[\protect\citeauthoryear{{Martini}, {Sivakoff}  \&
  {Mulchaey}}{{Martini} et~al.}{2009}]{Martini_2009}
{Martini} P.,  {Sivakoff} G.~R.,   {Mulchaey} J.~S.,  2009, \mn@doi [\apj]
  {10.1088/0004-637X/701/1/66}, 701, 66

\bibitem[\protect\citeauthoryear{{Martini} et~al.,}{{Martini}
  et~al.}{2013}]{Martini_2013}
{Martini} P.,  et~al., 2013, \mn@doi [\apj] {10.1088/0004-637X/768/1/1}, 768, 1

\bibitem[\protect\citeauthoryear{{Merloni} \& {Heinz}}{{Merloni} \&
  {Heinz}}{2013}]{Merloni_2013}
{Merloni} A.,  {Heinz} S.,  2013, {Evolution of Active Galactic Nuclei}.
p.~503, \mn@doi{10.1007/978-94-007-5609-0_11}

\bibitem[\protect\citeauthoryear{{Miyaji} et~al.,}{{Miyaji}
  et~al.}{2007}]{Miyaji_2007}
{Miyaji} T.,  et~al., 2007, \mn@doi [\apjs] {10.1086/516579}, 172, 396

\bibitem[\protect\citeauthoryear{{Moore}, {Katz}, {Lake}, {Dressler}  \&
  {Oemler}}{{Moore} et~al.}{1996}]{Moore_1996}
{Moore} B.,  {Katz} N.,  {Lake} G.,  {Dressler} A.,   {Oemler} A.,  1996,
  \mn@doi [\nat] {10.1038/379613a0}, 379, 613

\bibitem[\protect\citeauthoryear{{Papovich} et~al.,}{{Papovich}
  et~al.}{2010}]{Papovich_2010}
{Papovich} C.,  et~al., 2010, \mn@doi [\apj] {10.1088/0004-637X/716/2/1503},
  716, 1503

\bibitem[\protect\citeauthoryear{{Peng} et~al.,}{{Peng}
  et~al.}{2010}]{Peng_2010}
{Peng} Y.-j.,  et~al., 2010, \mn@doi [\apj] {10.1088/0004-637X/721/1/193}, 721,
  193

\bibitem[\protect\citeauthoryear{{Richstone}}{{Richstone}}{1976}]{Richstone_19%
76}
{Richstone} D.~O.,  1976, \mn@doi [\apj] {10.1086/154213}, 204, 642

\bibitem[\protect\citeauthoryear{{Saez} et~al.,}{{Saez}
  et~al.}{2015}]{Saez_2015}
{Saez} C.,  et~al., 2015, \mn@doi [\mnras] {10.1093/mnras/stv747}, 450, 2615

\bibitem[\protect\citeauthoryear{{Silverman} et~al.,}{{Silverman}
  et~al.}{2008}]{Silverman_2008}
{Silverman} J.~D.,  et~al., 2008, \mn@doi [\apj] {10.1086/529572}, 679, 118

\bibitem[\protect\citeauthoryear{{Simpson}, {Westoby}, {Arumugam}, {Ivison},
  {Hartley}  \& {Almaini}}{{Simpson} et~al.}{2013}]{Simpson_2013}
{Simpson} C.,  {Westoby} P.,  {Arumugam} V.,  {Ivison} R.,  {Hartley} W.,
  {Almaini} O.,  2013, \mn@doi [\mnras] {10.1093/mnras/stt940}, 433, 2647

\bibitem[\protect\citeauthoryear{{Springel}, {Di Matteo}  \&
  {Hernquist}}{{Springel} et~al.}{2005}]{Springel_2005}
{Springel} V.,  {Di Matteo} T.,   {Hernquist} L.,  2005, \mn@doi [\apjl]
  {10.1086/428772}, 620, L79

\bibitem[\protect\citeauthoryear{{Steidel}, {Adelberger}, {Shapley}, {Pettini},
  {Dickinson}  \& {Giavalisco}}{{Steidel} et~al.}{2003}]{Steidel_2003}
{Steidel} C.~C.,  {Adelberger} K.~L.,  {Shapley} A.~E.,  {Pettini} M.,
  {Dickinson} M.,   {Giavalisco} M.,  2003, \mn@doi [\apj] {10.1086/375772},
  592, 728

\bibitem[\protect\citeauthoryear{{Steidel}, {Shapley}, {Pettini}, {Adelberger},
  {Erb}, {Reddy}  \& {Hunt}}{{Steidel} et~al.}{2004}]{Steidel_2004}
{Steidel} C.~C.,  {Shapley} A.~E.,  {Pettini} M.,  {Adelberger} K.~L.,  {Erb}
  D.~K.,  {Reddy} N.~A.,   {Hunt} M.~P.,  2004, \mn@doi [\apj]
  {10.1086/381960}, 604, 534

\bibitem[\protect\citeauthoryear{{Tanaka}, {Finoguenov}  \& {Ueda}}{{Tanaka}
  et~al.}{2010}]{Tanaka_2010}
{Tanaka} M.,  {Finoguenov} A.,   {Ueda} Y.,  2010, \mn@doi [\apjl]
  {10.1088/2041-8205/716/2/L152}, 716, L152

\bibitem[\protect\citeauthoryear{{Tremaine} et~al.,}{{Tremaine}
  et~al.}{2002}]{Tremaine_2002}
{Tremaine} S.,  et~al., 2002, \mn@doi [\apj] {10.1086/341002}, 574, 740

\bibitem[\protect\citeauthoryear{Wilson}{Wilson}{1927}]{Wilson_1927}
Wilson E.~B.,  1927, \mn@doi [Journal of the American Statistical Association]
  {10.2307/2276774}, 22, 209

\bibitem[\protect\citeauthoryear{{van Breukelen} et~al.,}{{van Breukelen}
  et~al.}{2009}]{van_Breukelen_2009}
{van Breukelen} C.,  et~al., 2009, \mn@doi [\mnras]
  {10.1111/j.1365-2966.2009.14513.x}, 395, 11

\makeatother
\end{thebibliography}
\bsp

\label{lastpage}
\end{document}